\pdfoutput=1 % pour la compilation pdflatex par arXiv
\documentclass[doublecol,a4paper]{epl2} 

\usepackage[english,francais]{babel}

\usepackage{amsfonts,amsmath,amssymb,ifpdf,epsfig,graphicx}
\usepackage{mathrsfs} % POur \mathscr

\renewcommand{\leq}{\leqslant}
\renewcommand{\geq}{\geqslant}

\newcommand{\ket}[1]{|\kern.3ex#1\kern.3ex\rangle}
\newcommand{\bra}[1]{\langle\kern.3ex #1 \kern.3ex|}

\newcommand{\mean}[1]{\left\langle #1\right\rangle}
\newcommand{\smean}[1]{\langle #1\rangle}
\newcommand{\EXP}[1]{e^{#1}}         % exponentielle

\newcommand{\tr}[1]{\mathop{\mathrm{Tr}}\nolimits\left\{ #1 \right\}}  % Trace
\def\I{{\rm i}}

\newcommand{\derivp}[2]{\frac{\partial #1}{\partial #2}}

\def\D{{\rm d}}                  % la differenciation

\def\intpp{\smallsetminus\hspace{-0.38cm}\int}
\def\Ht{\tau_\mathrm{H}}
\def\Nc{N}

\def\Cgeo{C}
\def\rdc{R_\mathrm{dc}}
\def\lagZ{\mu_0}
\def\lagU{\mu_1}
\def\lagD{\mu_2}

\title{Capacitance and charge relaxation resistance of chaotic cavities -- Joint distribution of two linear statistics in the Laguerre ensemble of random matrices}

\shorttitle{Capacitance and charge relaxation resistance of chaotic cavities} %Insert here a short version of the title if it exceeds 70 characters

\author{Aur\'elien Grabsch\inst{1,2}\thanks{\email{aurelien.grabsch@ens-cachan.fr}} \and 
Christophe Texier\inst{2}\thanks{\email{christophe.texier@u-psud.fr}}}
\shortauthor{A. Grabsch and C. Texier}

\institute{                    
  \inst{1} \'{E}cole Normale Sup\'{e}rieure de Cachan %, 61 av. du Pr\'{e}sident Wilson, 
; 94235 Cachan cedex, France.\\
  \inst{2} Univ. Paris Sud ; CNRS ; Laboratoire de Physique Th\'eorique et Mod\`eles Statistiques, UMR 8626 ; 91405 Orsay cedex, France.}

\abstract{ 
 We consider the AC transport in a quantum RC circuit made of a coherent chaotic cavity with a top gate.
Within a random matrix approach, we study the joint distribution for the mesoscopic capacitance $C_\mu=(1/C+1/C_q)^{-1}$ and the charge relaxation resistance $R_q$, where $C$ is the geometric capacitance and $C_q$ the quantum capacitance.
We study the limit of a large number of conducting channels $N$ with a Coulomb gas method.
We obtain $\langle R_q\rangle\simeq h/(Ne^2)=R_\mathrm{dc}$ and show that the relative fluctuations are of order $1/N$ both for $C_q$ and $R_q$, with strong correlations 
$\langle \delta C_q\delta R_q\rangle/\sqrt{\langle \delta C_q^2\rangle\,\langle \delta R_q^2\rangle}\simeq+0.707$. The detailed analysis of large deviations involves a second order phase transition in the Coulomb gas. The two dimensional phase diagram is obtained. 
}

\begin{document}

\selectlanguage{english}

\maketitle

\section{Introduction}

\begin{figure}[!ht]
\centering
\includegraphics[width=0.35\textwidth]{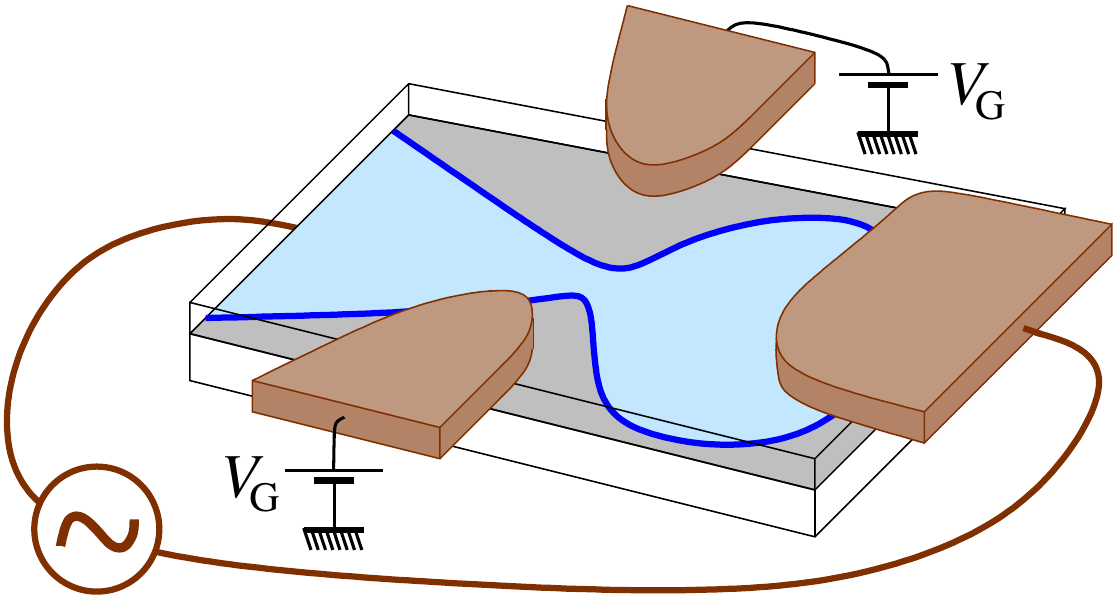}
\caption{A chaotic cavity in a 2DEG (blue) with top gates.}
\label{fig:QRCcircuit}
\end{figure}

The search for fast control and manipulation of charge in quantum coherent conductors has stimulated recent developments in time dependent response of mesoscopic structures~\cite{GabFevBerPlaCavEtiJinGla06}.
B\"uttiker, Pr\^etre and Thomas (BPT) \cite{ButPreTho93,ButThoPre93} have proposed a first theoretical description of the coherent AC response based on the scattering approach and a Thomas-Fermi treatment of electronic interactions.
When applied to the elementary RC circuit, BPT formalism provides the ``mesoscopic capacitance'' $C_\mu$ and the ``charge relaxation resistance'' $R_q$, two coefficients carrying information about the quantum dynamics of charges and involved in the impedance of the circuit,  $Z(\omega)=1/(-\I\omega\,C_\mu)+R_q+\mathcal{O}(\omega)$.
The capacitive response is splitted into two terms, $1/C_\mu=1/\Cgeo+1/C_q$, where $\Cgeo$ is the geometric capacitance deduced from the Poisson equation and $C_q$ the quantum capacitance (contribution of the density of states). 
In practice, the RC circuit can be realised with a chaotic cavity patterned in a two dimensional electron gas (2DEG) closed by a quantum point contact (QPC) and on the top of which is deposited a gate voltage capacitively coupled to the 2DEG (Fig.~\ref{fig:QRCcircuit}).
A remarkable feature that has attracted a lot of attention is the universal value $R_q=h/(2e^2)$, independent of the transmission properties of the contact, for a contact with a single spin-polarised channel. This prediction \cite{ButThoPre93} was demonstrated experimentally~\cite{GabFevBerPlaCavEtiJinGla06} by studying the circuit in the integer quantum Hall regime with filling factor one,
i.e. in an effectively one-dimensional situation.
This experimental advance has stimulated some theoretical efforts questioning the robustness of the universal result $R_q=h/(2e^2)$, shown to persist in the presence of Coulomb blockade, within a Hartree-Fock treatment of interaction 
\cite{NigLopBut06} %\cite{NigLopBut06,RinImrEnt08} 
or beyond mean-field 
\cite{HamJonKatMar10} %\cite{HamJonKatMar10,MorLeh10} 
% within Matveev's framework \cite{Mat95} 
(see also~\cite{EtzHorLed11}).
The transition between the universal \textit{half} quantum resistance and the QPC's DC resistance $R_q\simeq h/(e^2\mathcal{T})=\rdc$, where $\mathcal{T}$ is the transmission probability through the QPC, was shown to result from the presence of dephasing~\cite{NigBut08}.

If instead the system is studied in the weak magnetic field limit, electronic transport is sensitive to the complex (chaotic) dynamics inside the cavity.
When the dwell time is much larger than the Thouless time, random matrix theory (RMT) is a powerful approach, which has allowed to obtain several results: for a perfect contact with $\Nc$ conducting channels, the distribution of $C_q$ for $\Nc=1$ \cite{GopMelBut96}, $\Nc=2$ \cite{SavFyoSom01} and large $\Nc$ \cite{TexMaj13}.
The distribution of $R_q$ was found in the case $\Nc=2$ in Ref.~\cite{PedLanBut98} and 
the mean value $\mean{R_q}\simeq h/(e^2\Nc\mathcal{T})=\rdc$ for $\Nc\mathcal{T}\gg1$~\cite{But00}.~\footnote{
\label{footnoteBB}
The mean value of the admittance was first obtained in Ref.~\cite{BroBut97}, leading to 
  $\mean{R_q}=h/(\Nc e^2)\,\big[1+(2/\beta-1)/\Nc+\mathcal{O}(\Nc^{-2})\big]$, where the second term is the weak localisation correction (see also \cite{Bee97}). Only the fluctuations of the capacitive part of the response was obtained in this reference.
}
The statistical properties of $R_q$ for arbitrary $\Nc$ and the correlations with $C_q$ have remained unknown so far: it is the aim of this letter to answer this question.

Quantum mechanical properties of a single contact conductor (Fig.~\ref{fig:QRCcircuit}) with $\Nc$ conducting channels are encoded in the $\Nc\times\Nc$ scattering matrix $\mathcal{S}(\varepsilon)$.
Whereas many simple properties, like conductance or shot noise, can be obtained within the simple assumption of a uniform distribution of $\mathcal{S}$ over the unitary group~\cite{Bee97}, AC transport involves some information about the \textit{energy dependence} of the scattering matrix:
a Thomas-Fermi treatment of screening shows that the complex admittance $G(\omega)=1/Z(\omega)$ can be written as $1/G(\omega)=1/G_0(\omega)+1/(-\I\omega\Cgeo)$,
where $G_0(\omega)$ is the AC conductance of the non-interacting electron gas \cite{ButPreTho93}.
At zero temperature, its low frequency expansion takes the form
$G_0(\omega)=(e^2/h)\big[-\I\omega\tr{\mathcal{Q}}+(1/2)\omega^2\tr{\mathcal{Q}^2}+\mathcal{O}(\omega^3)\big]$
where 
$\mathcal{Q}=-\I\hbar\mathcal{S}^\dagger\partial\mathcal{S}/\partial\varepsilon$ is the Wigner-Smith time delay matrix \cite{CarNus02} taken at Fermi energy (finite temperature involves additional convolutions with Fermi functions); spin-degeneracy can be accounted for with the trace.
$C_\mu$ and $R_q$ appear in the $\omega\to0$ expansion of the admittance:
we get $C_q=(e^2/h)\tr{\mathcal{Q}}$, where $\nu\simeq\tr{\mathcal{Q}}/h$ is the density of states of the open conductor, and $R_q=[h/(2e^2)]\tr{\mathcal{Q}^2}/\big(\tr{\mathcal{Q}}\big)^2$~\cite{ButThoPre93}.
For a perfect contact such that $\mean{\mathcal{S}}=0$, where $\mean{\cdots}$ denotes ensemble averaging, the matrix $\mathcal{Q}^{-1}$ was shown to belong to the Laguerre ensemble of RMT~\cite{BroFraBee97,BroFraBee99}:
\begin{equation}
    \label{eq:Brouwer1997}
    \mathcal{P}(\gamma_1,\cdots,\gamma_{\Nc})
    \propto
    \prod_{i<j}|\gamma_i-\gamma_j|^\beta
    \prod_k \gamma_k^{\beta N/2}\EXP{-\beta\gamma_k/2}
    \:,
\end{equation}
where the proper time $\tau_i=\Ht/\gamma_i$ is an eigenvalue of $\mathcal{Q}$, $\Ht=h/\Delta$ is the Heisenberg time and $\Delta$ the mean level spacing.
$\beta\in\{1,\,2,\,4\}$ is the Dyson index corresponding to orthogonal, unitary or symplectic symmetry classes~\cite{Bee97}.
We introduce the dimensionless quantities
$s=\tr{\mathcal{Q}}/\Ht$ and $r=\Nc\tr{\mathcal{Q}^2}/\Ht^2$ in terms of which 
\begin{equation}
  C_q=\frac{e^2}{\Delta}\,s
  \hspace{0.5cm}\mbox{and} \hspace{0.5cm}
  R_q=\frac{h}{\Nc e^2} \, \frac{r}{2s^2}
  \:.
\end{equation}
Our aim is to analyse the joint distribution $P_\Nc(s,r)$.

\section{Main results} 

\begin{figure}[!ht]
\centering
\includegraphics[width=0.4\textwidth]{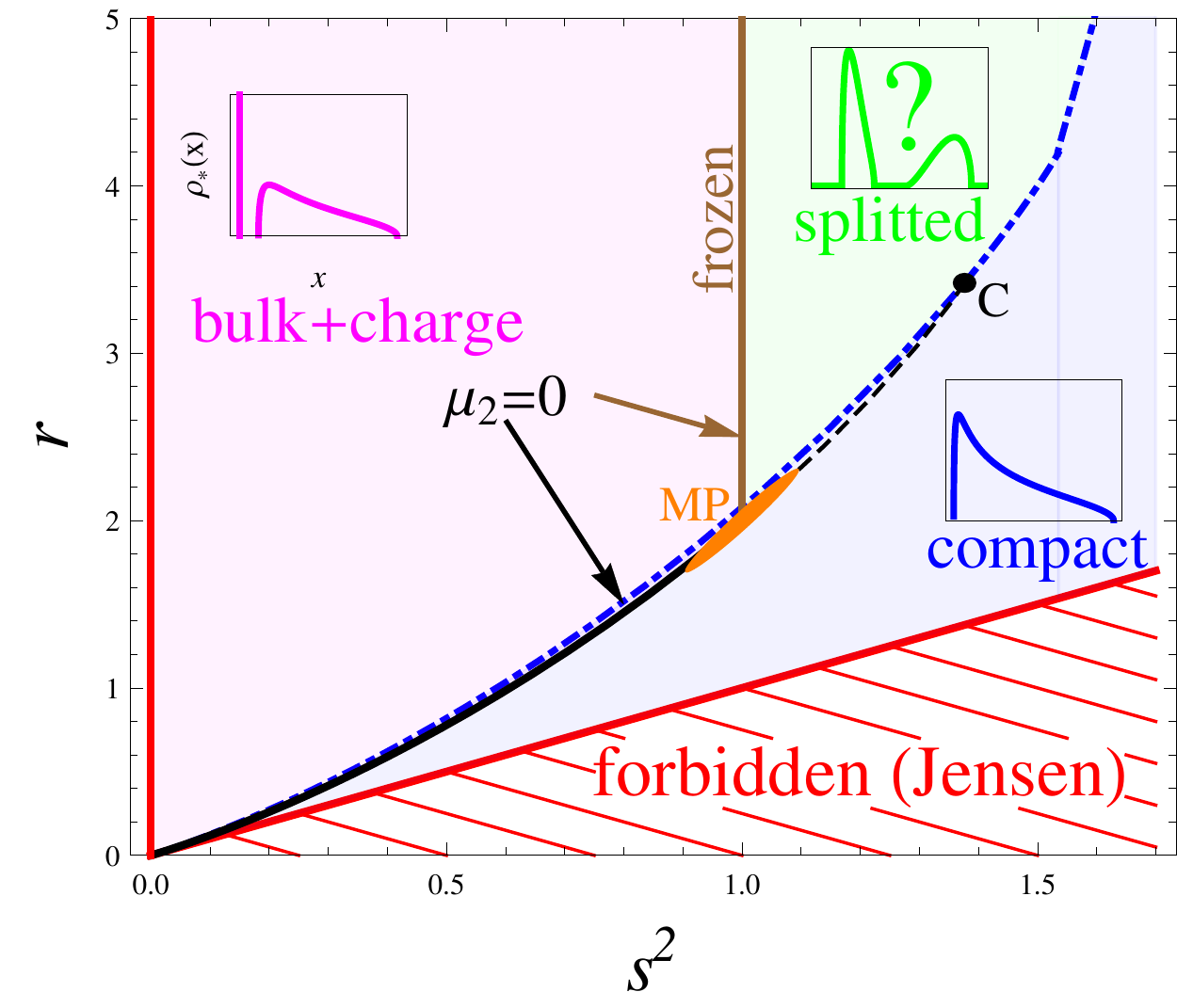}
\caption{Phase diagram of the Coulomb gas.
  The black line corresponds to a 2nd order transition. 
  Along the brown vertical line, the bulk density is frozen.
  The dotted-dashed blue line bounds the domain of existence of the compact phase. 
  }
\label{fig:PhaseDiagram}
\end{figure}

For large $\Nc$, it takes the scaling form
\begin{equation}
  \label{eq:DefLargeDeviation}
  P_\Nc(s,r) \underset{\Nc\to\infty}{\sim} 
  %\EXP{-(\beta/2)\Nc^2\Psi(s,r)}
  \exp\left\{ -(\beta/2)\Nc^2\Psi(s,r) \right\}
  \:,
\end{equation}
where $\Psi(s,r)$ is a large deviation function.
The distribution is dominated by a narrow Gaussian peak centered around $\mean{s}=1$ and $\mean{r}\simeq2$ (the orange ellipse marked MP on Fig.~\ref{fig:PhaseDiagram}), leading to $\mean{C_q}=e^2/\Delta$ and $\mean{R_q}\simeq h/(\Nc e^2)=\rdc$. We have recovered that $\mean{R_q}$ coincides with the DC resistance of the QPC, as expected~\cite{But00} (see footnote~\ref{footnoteBB}); although the scaling $R_q\sim1/\Nc$ was noticed in \cite{ButThoPre93}, we see that there is no simple combination rule as for the DC resistance since the contribution to $\mean{R_q}$ per channel crosses over from $h/(2e^2)$ for $\Nc=1$ to  $h/e^2$ for $\Nc\gg1$.~\footnote{for $\Nc=2$, using the distribution of Ref.~\cite{PedLanBut98}, we obtain $\mean{R_q}=(3/4)\,h/(2e^2)$ for $\beta=1$ and $\mean{R_q}=(5/7)\,h/(2e^2)$ for $\beta=2$.} 
Standard deviations are described by the expansion for $\delta s=s-1\ll1$ and $\delta r=r-2\ll1$:
\begin{equation}
  \label{eq:StandardDeviations}
  \Psi(s,r) \simeq
  \begin{pmatrix}
     \delta s & \delta r 
  \end{pmatrix} 
  \begin{pmatrix}
    4 & 24 \\ 24 & 160
  \end{pmatrix} ^{-1}
  \begin{pmatrix}
     \delta s \\ \delta r 
  \end{pmatrix}   
  \:,
\end{equation}
written in terms of the covariance matrix (this latter could also be obtained by other means~\cite{CunViv14,Viv14privcom}). %~\cite{CunViv14,Viv14privcom,Cun15}).
We recover 
$\mean{\delta C_q^2}/\mean{C_q}^2=\mathrm{Var}(s)\simeq4/(\beta\Nc^2)$~\cite{LehSavSokSom95,BroBut97,TexMaj13,MezSim13} and, writing $R_q\simeq[h/(\Nc e^2)](1-2\delta s+\delta r/2)$, we get
$\mean{\delta R_q^2}/\mean{R_q}^2\simeq8/(\beta\Nc^2)$. 
The strong correlations between proper times are responsible for much smaller relative fluctuations, $\sim1/\Nc$, than that of the sum of independent variables, $\sim1/\sqrt{\Nc}$. 
Eq.~\eqref{eq:StandardDeviations} shows that $s$ and $r$, and thus $C_q$ and $R_q$, are strongly correlated (this is illustrated by the orange ellipse on Fig.~\ref{fig:PhaseDiagram}):
$\mean{\delta C_q\delta R_q}/\sqrt{\smean{\delta C_q^2}\smean{\delta R_q^2}}=+1/\sqrt{2}$.
The $RC$ time $\tau_{RC}=R_qC_\mu = [r/(2s)]\,\tau_\mathrm{dwell}/(1+s\,E_C/\Delta)$ is also of interest, where $\tau_\mathrm{dwell}=h/(\Nc\Delta)$ is the dwell time and $E_C=e^2/C$ the charging energy.
We find $\mean{\tau_{RC}}\simeq\tau_\mathrm{dwell}/(1+E_C/\Delta)$, with relative fluctuations of order $1/\Nc$ as well.

The large deviation function $\Psi(s,r)$ has a rich structure whose analysis can be formulated as the study of thermodynamic properties of a one-dimensional gas of $\Nc$ particles with logarithmic interactions (the \og Coulomb gas \fg{}).
The available region in the plane $(s,r)$ is bounded by two constraints:
$r\geq s^2$ coming from the Jensen's inequality and $r\leq\Nc s^2$ (vertical axis $s=0$ in the thermodynamic limit $\Nc\to\infty$) due to the positivity of the $\gamma_i$'s, implying $h/(2\Nc e^2)\leq R_q\leq h/(2e^2)$.
In the right part of the phase diagram (Fig.~\ref{fig:PhaseDiagram}), 
we have found a particle density with compact support. Near the lower boundary, we get a logarithmic divergence of the large deviation function (i.e. the energy of the gas) 
\begin{equation}
  \label{eq:LargeDevNearJensen}
  \Psi(s,r) \simeq
  -\frac{1}{2}\ln(r-s^2)
  \hspace{0.25cm}\mbox{for }
  r-s^2\to0^+
  \:
\end{equation}
 (cf. Fig~\ref{fig:PsiSR}).
This behaviour  corresponds to the rapid vanishing $P_\Nc(s,r)\sim(r-s^2)^{\beta\Nc^2/4}$.
Correspondingly, the distribution of $R_q$ vanishes as $\sim[R_q-h/(2\Nc e^2)]^{\beta\Nc^2/4}$ for $R_q\to h/(2\Nc e^2)$.
Across the black line of Fig.~\ref{fig:PhaseDiagram}, the Coulomb gas exhibits a phase transition towards a phase characterised by a density with a non compact support, where one particle splits off the bulk and brings a ``macroscopic''  contribution to $\tr{\mathcal{Q}^2}$.
In this region the large deviation function is independent of~$r$: 
\begin{equation}
  \label{eq:LargeDevNonCompactPhase}
  \Psi(s,r) \simeq \Phi_-(s)  
\end{equation} 
where $\Phi_-(s)$ is the large deviation function related to the marginal law of the variable $s$ obtained in Ref.~\cite{TexMaj13}, with limiting behaviours 
$\Phi_-(s)\simeq(1/4)(s-1)^2$ for $s\to1^-$ and 
$\Phi_-(s)\simeq1/s+(3/2)\ln s
$ 
for $s\ll1$  (blue curve on Fig~\ref{fig:PsiSR}).

\section{Coulomb gas method}

The Coulomb gas method, introduced by Dyson, %~\cite{Dys62a,Dys62b,Dys62c} 
starts with the interpretation of the distribution~\eqref{eq:StandardDeviations} as the Gibbs equilibrium measure for $\Nc$ ``charges'' of ``positions'' $\gamma_i$'s
\cite{DeaMaj06,DeaMaj08,MajVer09,VivMajBoh10,TexMaj13,For10}.
%\cite{DeaMaj06,DeaMaj08,MajVer09,VivMajBoh10,NadMajVer10,NadMajVer11,TexMaj13,For10}.
Rescaling the dimensionless rates as $\gamma_i=\Nc x_i$ and introducing the charge density $\rho(x)=(1/\Nc)\sum_{i=1}^{\Nc}\delta(x-x_i)$ lead to $\mathcal{P}(\gamma_1,\cdots,\gamma_{\Nc})\propto\exp\big\{-(\beta/2)\Nc^2\mathscr{E}[\rho]\big\}$, where the energy
\begin{align}
  \label{eq:Energy}
    \mathscr{E}[\rho] =& \int_0^\infty\D x\,\rho(x)\,(x-\ln x)
    \nonumber\\
    &-\int_0^\infty\D x\D x'\,\rho(x)\,\rho(x')\,\ln|x-x'|
\end{align}
describes confinement by the potential $V(x)=x-\ln x$ and logarithmic repulsion between the charges.
The analysis of the joint distribution of $s=(1/\Nc)\sum_ix_i^{-1}$ and $r=(1/\Nc)\sum_ix_i^{-2}$ can be recast as an energy minimization problem under three constraints, leading to consider the ``free energy''
$
\mathscr{F}[\rho] = \mathscr{E}[\rho] 
+ \lagZ \, \big( \int\D x\,\rho(x)-1 \big)
+ \lagU \, \big( \int\D x\,\rho(x)/x   - s \big)
+ \lagD \, \big( \int\D x\,\rho(x)/x^2 - r \big)
$,
where $\lagZ$, $\lagU$ and $\lagD$ are three Lagrange multipliers.
Minimization of the free energy, $\delta\mathscr{F}[\rho]/\delta\rho(x)=0$, and derivation with respect to $x$, leads to the equilibrium condition
\begin{equation}
  \label{eq:SD}
   1 - \frac{1}{x} - \frac{\lagU}{x^2} - \frac{2\lagD}{x^3} 
  =2\intpp_a^b\D x'\,\frac{\rho(x')}{x-x'}
  \hspace{0.25cm}\mbox{for }
  x\in[a,b]
\end{equation}
where $\smallsetminus\hspace{-0.3cm}\int$ represents the principal part. 
We have assumed that the distribution has a compact support $[a,b]$.
Eq.~\eqref{eq:SD} expresses equilibration of the two forces (confinement and repulsion) acting on the charge at $x$. 
Solving Eq.~\eqref{eq:SD} provides the density as a function of $\lagU$ and $\lagD$; imposing the three constraints furnishes the dependence of the Lagrange multipliers in $s$ and $r$, i.e. two functions $\lagU=\lagU^\star(s,r)$ and $\lagD=\lagD^\star(s,r)$, from which we deduce the equilibrium density  $\rho_\star(x;s,r)$, as a function of $s$ and~$r$.
Denoting by $\rho_\mathrm{MP}(x)$ the solution for $\lagU=\lagD=0$, we get the form \eqref{eq:DefLargeDeviation} with $\Psi(s,r)=\mathscr{E}[\rho_\star]-\mathscr{E}[\rho_\mathrm{MP}]$, where the second term comes from the normalisation of $P_\Nc(s,r)$.

\section{Thermodynamic identities}

We have obtained~\cite{GraMajTex14} 
\begin{align}
  \label{eq:ThermoIdentities}
  \derivp{\mathscr{E}[\rho_\star]}{s} = -\lagU^\star(s,r)  
  \hspace{0.5cm}\mbox{and} \hspace{0.5cm}
  \derivp{\mathscr{E}[\rho_\star]}{r} = -\lagD^\star(s,r)  
  \:,
\end{align}
which have simplified the determination of the energy:
knowing $\rho_\star$, instead of a direct calculation  of $\mathscr{E}[\rho_\star]$ from Eq.~\eqref{eq:Energy}, as in  %Refs.~\cite{DeaMaj06,DeaMaj08,MajVer09,VivMajBoh10,NadMajVer10,NadMajVer11,TexMaj13}, 
Refs.~\cite{DeaMaj06,DeaMaj08,MajVer09,VivMajBoh10,TexMaj13}, 
the identities \eqref{eq:ThermoIdentities} allow to obtain $\mathscr{E}[\rho_\star]$ more efficiently by integration of the Lagrange multipliers, which solve algebraic equations obtained below.

\section{Compact phase}

Eq.~\eqref{eq:SD} is solved with Tricomi's theorem~\cite{Tri57} (see also Ref.~\cite{DeaMaj08,VivMajBoh10}). We get 
\begin{equation}
  \label{eq:RhoStar}
  \rho_\star(x;s,r) = \frac{x^2+c\,x+d}{2\pi x^3}\sqrt{(b-x)(x-a)}
  \:,
\end{equation}
where
$c=\lagU/\sqrt{ab}+\lagD\,(a+b)/(ab)^{3/2}$ and $d=2\lagD/\sqrt{ab}$. 
% (the solution obtained in \cite{TexMaj13} is recovered by setting $\lagD=0$).
Given $s$ and $r$, we find the support $[a,b]$ by solving 
%\begin{widetext}
\begin{align}
  \label{eq:EqForS}
  s &= 
    \Big[2u\,(3-4u+18u^2-4u^3+3u^4)
    \nonumber \\ & \hspace{1cm} - v\,(1-u)^4(1+4u+u^2)\Big]
  \frac{1}{32u^3v} 
  %%  \frac{2u\,(3-4u+18u^2-4u^3+3u^4)- v\,(1-u)^4(1+4u+u^2)}{32u^3v} 
  \:,
  \\
  \label{eq:EqForR}
  r &= 
    \Big[ 2u\,(9-10u+39u^2-12u^3+39u^4-10u^5+9u^6)
    \nonumber \\ & \hspace{1cm}  -3v\,(1-u^2)^4 \Big]
    \frac{1}{128u^4v^2}
  %%   \frac{2u\,(9-10u+39u^2-12u^3+39u^4-10u^5+9u^6)-3v\,(1-u^2)^4}{128u^4v^2}
  \:,
\end{align}
where $u=\sqrt{a/b}$ and $v=\sqrt{ab}$.
Then the Lagrange multipliers are given by 
\begin{align}
  \label{eq:Mu1}
  \lagU &= 
  \Big[ 2u\,(-9+4u-6u^2+4u^3-9u^4)
    \nonumber \\ & \hspace{1cm}  +3v\,(1-u^2)^2(1+u^2) \Big]
    \frac{v}{2u(1-u^2)^2} 
  %%  v\,\frac{2u\,(-9+4u-6u^2+4u^3-9u^4)+3v\,(1-u^2)^2(1+u^2)}{2u(1-u^2)^2} 
  \:,
  \\
  \label{eq:Mu2}
  \lagD &= 
   -v^2\,\frac{2u\,(-3+2u-3u^2)+v\,(1-u^2)^2}{(1-u^2)^2} 
  \:.
\end{align}
%\end{widetext}

Setting $\lagU=\lagD=0$ corresponds to relaxing the two constraints and to searching for the global minimum of $\mathscr{E}[\rho_\star]$ [i.e. the maximum of $P_\Nc(s,r)$]. This leads to $u=x_-\equiv u_\mathrm{MP}$ and $v=1\equiv v_\mathrm{MP}$, where $x_\pm=3\pm2\sqrt2$, corresponding to the Mar\v{c}enko-Pastur (MP) law \cite{MarPas67,BroFraBee99,TexMaj13}
$\rho_\mathrm{MP}(x)=\sqrt{(x-x_-)(x_+-x)}/(2\pi x)$.
We deduce the two typical values $s=\int\D x\,\rho_\mathrm{MP}(x)/x=1$ and $r=\int\D x\,\rho_\mathrm{MP}(x)/x^2=2$, i.e. $\rho_\star(x;1,2)=\rho_\mathrm{MP}(x)$.
%, corresponding to $\mean{C_q}=e^2/\Delta$ and $\mean{R_q}\simeq h/(\Nc e^2)$ aforementioned.

Small deviations around this solution can be easily studied by using \eqref{eq:ThermoIdentities}, leading to
$
\mathscr{E}[\rho_\star]
\simeq\mathscr{E}[\rho_\mathrm{MP}]
-(1/2)(s-1)^2\partial\lagU/\partial s\big|_\mathrm{MP}
-(1/2)(r-2)^2\partial\lagD/\partial r\big|_\mathrm{MP}
-(s-1)(r-2)\partial\lagU/\partial r\big|_\mathrm{MP} 
$,
where derivatives are calculated at $u=u_\mathrm{MP}$ and $v=v_\mathrm{MP}$, hence Eq.~\eqref{eq:StandardDeviations}.

The vicinity of the lowest boundary $r-s^2\to0^+$ of the phase diagram, i.e. $u\to1^-$, is studied as follows: Eqs.~(\ref{eq:EqForS},\ref{eq:EqForR}) give $u\simeq1-2\sqrt{r-s^2}/s$ and $v\simeq1/s$. Then the expansion of the Lagrange multipliers (\ref{eq:Mu1},\ref{eq:Mu2}) and their integration thanks to \eqref{eq:ThermoIdentities} eventually lead to~\eqref{eq:LargeDevNearJensen}.

These results have been confirmed by Monte Carlo simulations of the Coulomb gas which will be described elsewhere~\cite{GraMajTex14}.

\section{Phase transition}

Besides the two natural constraints $r\geq s^2$ and $r\leq\Nc s^2$, Eqs.~(\ref{eq:EqForS},\ref{eq:EqForR}) do not have real solutions in the full available domain.
A necessary condition for the existence of the solution \eqref{eq:RhoStar} is $x^2+c\,x+d\geq0$ for $x\in[a,b]$, which leads to the domain bounded by the blue dashed-dotted line on Fig.~\ref{fig:PhaseDiagram}.
A similar phenomenon has already occured elsewhere \cite{MajVer09,MajSch14,TexMaj13}. 
%\cite{MajVer09,MajSch14,NadMajVer10,NadMajVer11,TexMaj13}. 
%, like the study of the largest eigenvalue of random matrices \cite{MajVer09,MajSch14}, large deviations of the entanglement entropy~\cite{NadMajVer10,NadMajVer11} or Wigner time delay \cite{TexMaj13}.
The origin of the problem can be identified by inspection of the effective confining potential 
$V_\mathrm{eff}(x)%=V(x)+\lagU/x+\lagD/x^2
=x-\ln x+\lagU/x+\lagD/x^2$ 
from which the force of the left hand side of Eq.~\eqref{eq:SD} derives: 
when $\lagD<0$ the compact phase is unstable (or metastable). The line $\lagD=0$ thus defines where a phase transition takes place. 

\begin{figure}[!ht]
\centering
\includegraphics[width=0.4\textwidth]{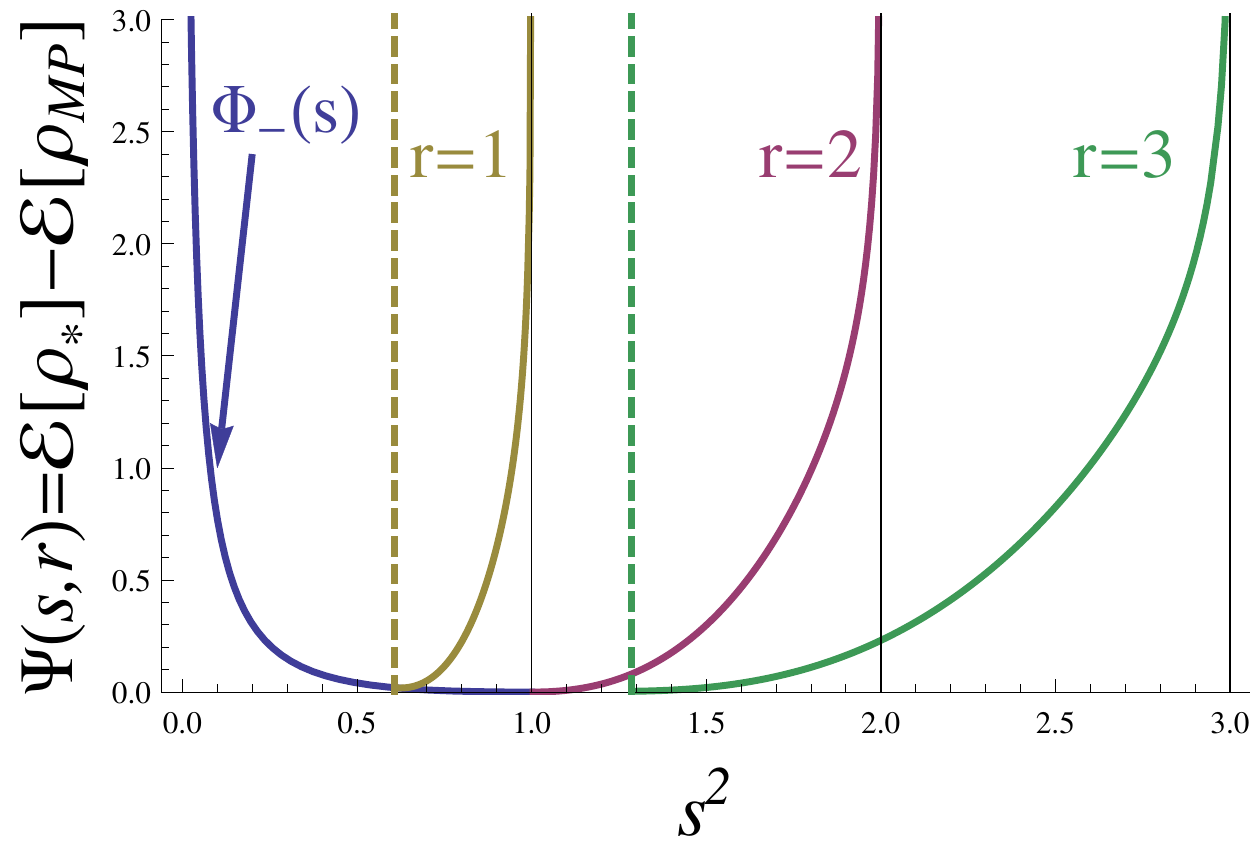}
\caption{
Large deviation function for three values of $r$. Dashed lines set the boundary for the compact phase.
}
\label{fig:PsiSR}
\end{figure}

In the region of the phase diagram at the left of the line $\lagD=0$ (Fig.~\ref{fig:PhaseDiagram}), the compact phase (density with compact support) is energetically unfavorable or unstable.
The new phase corresponds to one charge at $x_1$ splitted off the bulk:
$\rho(x)=(1/\Nc)\,\delta(x-x_1)+\tilde{\rho}(x)$ where the density $\tilde{\rho}(x)$ with compact support describes the remaining $\Nc-1$ charges forming the ``bulk''. 
The fact that one proper time $\tau_1=\Ht/(\Nc x_1)$ is dominant can be interpreted as the contribution of a narrow resonance in the original scattering problem~\cite{TexMaj13}.
The bulk density $\tilde{\rho}(x)$ and the position $x_1$ are determined by 
% the equilibrium condition for a charge at $x$ in the bulk 
$\delta\mathscr{F}[\rho]/\delta\tilde{\rho}(x)=0$ and 
%for the isolated charge 
$\partial\mathscr{F}[\rho]/\partial x_1=0$, leading to
\begin{align}
  \label{eq:EquilBulkCharge}
   1 - \frac{1}{x} - \frac{\lagU}{x^2} - \frac{2\lagD}{x^3}
   &=\frac{2/\Nc}{x-x_1} 
  +2\intpp_a^b\D x'\,\frac{\tilde\rho(x')}{x-x'}
  \\
  \label{eq:EquilIsolatedCharge}
   1 - \frac{1}{x_1} - \frac{\lagU}{x_1^2} - \frac{2\lagD}{x_1^3} 
   &=2\int_a^b\D x'\,\frac{\tilde\rho(x')}{x_1-x'}
  \:.
\end{align}
We expect $x_1\to0$ while the bulk remains at distance $a\gg x_1$ from the origin. The analysis of \eqref{eq:EquilIsolatedCharge} shows that $\lagD$ must absorb the main divergence of the left hand side in the $x_1\to0$ limit; thus $2\lagD\simeq-\lagU\,x_1-x_1^2\to0$. 
It follows that, if $\Nc\gg1$, Eq.~\eqref{eq:EquilBulkCharge} has for solution the one describing the marginal law of the variable $s$ studied in Ref.~\cite{TexMaj13}, leading to an energy profile flat in the $r$ direction (at lowest order in $1/\Nc$), Eq.~\eqref{eq:LargeDevNonCompactPhase} (cf. Fig~\ref{fig:PsiSR}).
From $r=1/(\Nc x_1^2)+\int\D x\,\tilde\rho(x)/x^2$,
we deduce $x_1\simeq1/\sqrt{\Nc(r-r(s))}$ where $r=r(s)$ is the line where $\lagD=0$ (black line on Fig.~\ref{fig:PhaseDiagram}).
The energy of the gas contains a contribution $-(1/\Nc)\ln x_1$ from the potential energy of the isolated charge,
subdominant in $1/\Nc$ but diverging as $x_1\to0$.~\footnote{
  Note that the energy contains other $1/\Nc$ terms and the entropy of the bulk density \cite{DeaMaj08}, neglected here, also produces another $s$-dependent contribution of order $1/\Nc$ to be added to the energy~$\mathscr{E}[\tilde{\rho}]$.
}
Hence $P_\Nc(s,r)\sim(r-r(s))^{-\beta\Nc/4}\exp\big\{-(\beta/2)\Nc^2\Phi_-(s)\big\}$. 

For $r\leq2$ and $s\leq1$, the phase transition takes place at the black line of Fig.~\ref{fig:PhaseDiagram} (the dashed part between MP and C corresponds to a metastable branch for the compact phase~\cite{TexMaj13});  $\lagU$ and $\lagD$ are continuous across the line but $\lagD$ is not differentiable, hence the phase transition is \textit{second order}.
For $s\geq1$, the line $\lagD=0$ corresponds to the frozen phase of Ref.~\cite{TexMaj13} (with $\lagU\simeq0$): $\tilde\rho(x)=\rho_\mathrm{MP}(x)+\mathcal{O}(\Nc^{-1})$ with 
%an isolated charge at
$x_1\simeq1/[\Nc(s-1)]$. 
This frozen phase exists on the line $r=1/(\Nc x_1^2)+\int\D x\,\tilde\rho(x)/x^2\simeq\Nc(s-1)^2+2$
(brown line of Fig.~\ref{fig:PhaseDiagram}).
Although we have not yet determined $\rho_\star$ in the last region between the brown vertical line and the blue dashed-dotted line, nor its energy, we have strong reasons to believe that the support of the density is splitted into two disconnected intervals, like it occurs in other problems (e.g.~\cite{VivMajBoh10,MajNadScaViv09}); this is supported by the analysis of the shape of the effective potential $V_\mathrm{eff}(x)$ (when $\lagD>0$ and $\lagU<0$) and the study of the polynomial $x^2+c\,x+d$ in \eqref{eq:RhoStar} in the vicinity of the dotted-dashed blue line~\cite{GraMajTex14}. 

Recently, the existence of phase transitions driven by large deviations was studied in Ref.~\cite{SzaEvaMaj14} by considering two linear statistics of the form $s=\sum_iy_i$ and $r=\sum_iy_i^\eta$ where the random variables $y_i$ are identical and \textit{uncorrelated} (in our case, the marginal law of $\tau_i$ has a power law tail).
Although part of the phase diagram is reminiscent of ours, we see that correlations induce a richer structure.

\section{Conclusion}

We have studied the joint distribution for two linear statistics in the Laguerre ensemble of RMT, leading to the study of a two dimensional phase diagram for the Coulomb gas.
Our analysis has been greatly simplified by the use of the thermodynamics identities \eqref{eq:ThermoIdentities}.
We have obtained the large deviation functions and the phase diagram; the region remained unexplored here will be hopefully unveiled in a forthcoming paper~\cite{GraMajTex14}. 
We have deduced some properties for the distribution of the charge relaxation resistance: position and width of the dominant Gaussian peak, behaviour near the lower boundary, for $R_q\to h/(2\Nc e^2)$. However the large deviations related to the behaviour of the distribution near the upper boundary, for $R_q\to h/(2e^2)$, is still unknown.
The universal value of the charge relaxation resistance for $N=1$ channel has been measured for a quantum dot in the integer quantum Hall regime~\cite{GabFevBerPlaCavEtiJinGla06}~;
similar experimental study in the weak magnetic field regime with many open channels should allow for an investigation of the charge relaxation resistance's statistical properties, hopefully revealing the rich structure of its distribution in connection with the phase transitions in the Coulomb gas discussed here.
Such a study relies on the possibility to realize sample averaging, which was recently achieved for a single sample by using gate voltages in an appropriate way in the remarkable experiment on phase coherence time~\cite{RauAmaGroPotShtGol12}.
 
% Our study opens many challenging issues:
%(\textit{i}) an interesting question would be to extend our analysis to the case of tunnel contacts, as for the marginal laws of proper times \cite{FyoSavSom97}.
%(\textit{ii}) BPT formalism allows to consider multiterminal conductors: the study of out-of-equilibrium situations would lead naturally to consider quantities similar to $C_q$ and $R_q$ involving partial sums over channels.
%(\textit{iii})
Another challenging issue is related to the possible decorrelation of linear statistics, as pointed out in the covariance analysis of the conductance $g$ and the shot noise $p$ of a two terminal conductor with $N_1$ and $N_2$ channels at the two contacts~\cite{SavSomWie08,CunViv14}. 
%in particular for the case of the conductance and the shot noise of a two terminal conductor, with $N_1$ and $N_2$ channels at the two contacts, the decorrelation is driven by the difference $\alpha=|N_1-N_2|$ and occurs for symmetric contacts ($\alpha=0$). 
The vanishing of the covariance obtained in~\cite{SavSomWie08}, $\mathrm{Cov}(g,p)\simeq(2/\beta)\big[(1-2/\beta)^2-(N_1-N_2)^2\big](N_1N_2)^2/(N_1+N_2)^6$, when  $N_1=N_2$ and $\beta=2$, was attributed to the symmetry $T_i\leftrightarrow1-T_i$ of the distribution of the transmission probabilities $T_i$'s. 
In the highly asymmetric limit, $N_1\gg N_2$, the results of Ref.~\cite{SavSomWie08} lead to a full anticorrelation 
$\mathrm{Cov}(g,p)/\sqrt{\mathrm{Var}(g)\mathrm{Var}(p)}\simeq-1$.
An interesting question would be to examine these phenomena at the level of the joint distribution itself and the implication for the thermodynamics of the Coulomb gas.

\section{Acknowledgements}

We acknowledge stimulating discussions with Satya Majumdar, Christophe Mora, Gr\'egory Schehr, Denis Ullmo and Pierpaolo Vivo.

%\bibliographystyle{eplbib}
%\bibliography{biblio}

\end{document}